\documentclass[final]{aipproc}
\usepackage{psfrag,graphicx,amsmath}
\layoutstyle{6x9}

\def\lesssim{\mathrel{\hbox{\rlap{\hbox{\lower4pt\hbox{$\sim$}}}\hbox{$<$}}}}
\def\gtrsim{\mathrel{\hbox{\rlap{\hbox{\lower4pt\hbox{$\sim$}}}\hbox{$>$}}}}

\begin{document}

\title{Turbulence and Mixing in the Intracluster Medium}


\author{P. Sharma}{
 address={
Theoretical Astrophysics Center and Astronomy Department, University of California, Berkeley, CA 94720, USA}
}

\author{B. D. G. Chandran }{
 address={ 
Space Science Center and Department of Physics, University of New Hampshire, Durham, NH 03824, USA}
}

\author{E. Quataert}{
address={
Theoretical Astrophysics Center and Astronomy Department, University of California, Berkeley, CA 94720, USA} 
}

\author{I. J. Parrish}{
address={
Theoretical Astrophysics Center and Astronomy Department, University of California, Berkeley, CA 94720, USA}
}

\begin{abstract}
  The intracluster medium (ICM) is stably stratified in the
  hydrodynamic sense with the entropy~$s$ increasing outwards.
  However, thermal conduction along magnetic field lines fundamentally
  changes the stability of the ICM, leading to the ``heat-flux
  buoyancy instability'' when $dT/dr>0$ and the ``magnetothermal
  instability'' when $dT/dr<0$. The ICM is thus buoyantly unstable
  regardless of the signs of $dT/dr$ and $ds/dr$.  On the other hand,
  these temperature-gradient-driven instabilities saturate by
  reorienting the magnetic field (perpendicular to $\hat{\bf r}$ when
  $dT/dr>0$ and parallel to $\hat{\bf r}$ when $dT/dr<0$), without
  generating sustained convection.  We show that after an
  anisotropically conducting plasma reaches this nonlinearly stable magnetic
  configuration, it experiences a buoyant restoring force that resists
  further distortions of the magnetic field.  This restoring force is
  analogous to the buoyant restoring force experienced by a stably
  stratified adiabatic plasma.  We argue that in order for a driving
  mechanism (e.g, galaxy motions or cosmic-ray buoyancy) to overcome
  this restoring force and generate turbulence in the ICM, the
  strength of the driving must exceed a threshold,
  corresponding to turbulent velocities $\gtrsim 10 -100
  \;\mbox{km/s}$. For weaker driving, the ICM remains in its
  nonlinearly stable magnetic configuration, and turbulent mixing is
  effectively absent. We discuss the implications of these findings
  for the turbulent diffusion of metals and heat in the~ICM.

\end{abstract}

\maketitle

\section{Introduction}

One of the important developments in the study of galaxy clusters during the last decade has been the discovery that very little gas cools to low temperatures 
in cluster cores \cite{pet03,ode08}. Strong cooling flows are expected in the absence of non-adiabatic heating of the ICM \cite{fab94}. The lack of cooler gas suggests that heating, most likely due to feedback from the 
central Active Galactic Nucleus (AGN), roughly balances cooling. Although the plasma does not cool catastrophically, observational signatures of cooling (H$\alpha$ filaments) and feedback (radio bubbles) are seen 
in many clusters \cite{cav08}. Energetically, there are many heat sources that can balance cooling in cluster cores:
thermal conduction from larger radii \cite{ber86}; heating by jets and bubbles \cite{bin95}; heating by cosmic rays \cite{cha07,guo08}; etc. However, the challenge is to identify an {\em isotropic} heating mechanism that can balance cooling over many cooling timescales without violating the observational constraints.

The focus of this paper is on the transport and mixing properties of the ICM, and not on a particular ICM heating mechanism. 
Important advances in understanding the convective stability of the ICM have been made over the past few years. In particular, it has been realized that the ICM is buoyantly unstable in the presence of thermal conduction along magnetic field lines, regardless of the signs of the temperature and entropy gradients, $dT/dr$ and $ds/dr$. In cluster cores, where $dT/dr>0$, the ICM is unstable to the ``heat-flux-driven buoyancy instability'' (HBI) \cite{qua08}, and at larger radii where $dT/dr<0$ the plasma is unstable to the ``magnetothermal instability'' (MTI) \cite{bal00}.  The action of these instabilities is to reorient the magnetic field, so that field lines become perpendicular to~$\nabla T$ when $dT/dr>0$ and parallel to $\nabla T$ when $dT/dr<0$ (\cite{par07,par08,sha08}). In cluster cores, the reorientation of the magnetic field effectively shuts off thermal conduction, resulting in a catastrophic cooling flow \cite{par09}.  Our goal in this paper is to evaluate the broader effect of thermal conduction on the mixing and transport properties of the ICM; in particular, to determine how effectively heating concentrated in some specific volume due to external driving (e.g., by AGN jets, galaxy wakes) can be redistributed throughout the ICM. To address this we focus on convective transport driven by cosmic rays, and show that the turbulence is spread out (both in $r$ and in $\theta$ and $\phi$) more effectively with thermal conduction than without it. We interpret this in terms of the Richardson number, the ratio of the buoyancy and turbulent forces. The governing equations, numerical simulations, and the initial set-up have been discussed in detail in \cite{sha09} and have not been repeated here.

\section{Results}
The numerical simulations that we discuss are based on the standard MHD
equations but with thermal conduction (both isotropic and anisotropic [i.e.,
along magnetic field lines]). In addition, cosmic rays are evolved as a
relativistic fluid with the same bulk velocity as that of the thermal plasma,
which couples to the plasma via its pressure in the momentum equation.
Diffusion of cosmic rays along magnetic field lines is included; however, cosmic
rays, within a reasonable range of diffusion coefficients, are effectively
adiabatic for the scales of interest.

In \cite{sha09} we studied in detail the convective instability driven by the
gradient in the cosmic-ray `entropy' (defined as $p_{\rm cr}/\rho^{4/3}$, where
$p_{\rm cr}$ is the cosmic-ray pressure and $\rho$ is the plasma density), which
arises when the cosmic-ray pressure is a significant fraction of the thermal
pressure and the cosmic-ray `entropy' decreases outwards. We found that
cosmic-ray-driven convection stirs the ICM, both in radius and angle, when the
cosmic-ray pressure is sufficiently large.

Here, we focus on the general mixing properties of the ICM; cosmic-ray
convection is only one of the plausible mechanisms for generating turbulence
(others being, AGN jets, galactic wakes, etc.). We inject cosmic rays only
within a narrow range of angles ($\theta<30^0$) about the poles; the cosmic-ray
pressure is built up by a steep (such that the cosmic-ray `entropy' decreases
outwards) cosmic-ray source term (see \cite{sha09} for details).  We compare
three simulations with initial temperature and density profiles similar to those
observed in typical cluster cores (the simulation parameters are the same as in
runs CR30 and CR30-ad of \cite{sha09}, except that these are axisymmetric runs):
(1) plasma with thermal conduction (Spitzer value) along magnetic field lines;
(2) plasma with isotropic (Spitzer value) thermal conduction; and (3) adiabatic plasma
with no thermal conduction. Plasma cooling is not included in these runs, as the
focus is on the effect of thermal conduction on plasma mixing.

\begin{figure}
\includegraphics[width=4in,height=4in]{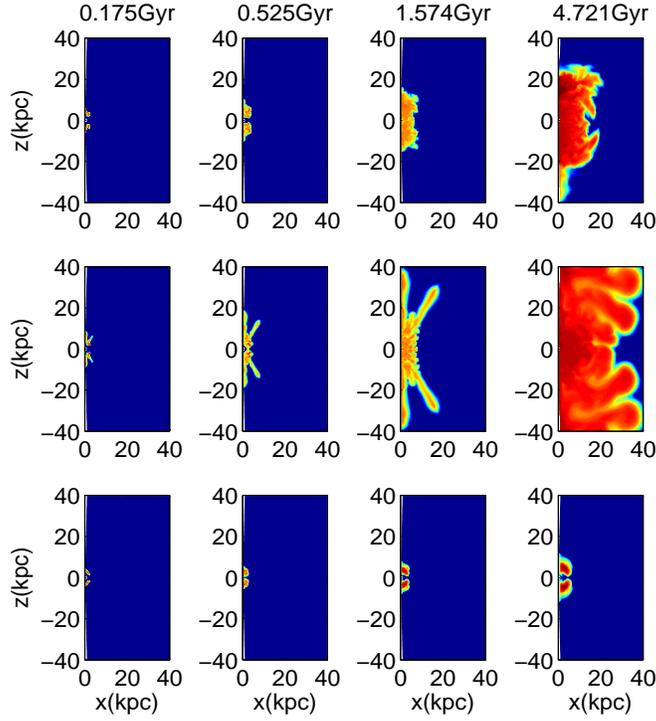}
\caption{Metallicity (logarithmic) profiles (scaled by its maximum value in the particular plot) at different times for runs with anisotropic thermal conduction (top row),
isotropic thermal conduction (middle row), and adiabatic plasma (bottom row). A passive scalar
initialized within inner four grid points (<1.25 kpc) is used as a proxy for metallicity. While mixing
driven by cosmic-ray convection is most efficient with isotropic thermal conduction, it is least effective for an adiabatic plasma. See the text for details.}\label{fig}
\end{figure}

Figure (\ref{fig}) shows metallicity profiles at different times for runs with anisotropic conduction (top row), isotropic conduction (middle row), and no thermal conduction (i.e., adiabatic plasma; bottom row). Mixing is efficient with thermal conduction, and a large volume of plasma (even at the equator, although the cosmic-ray source only operates near the poles) is mixed by cosmic-ray-driven convection. This figure is
analogous to Figure (11) in \cite{sha09}, but is based on two-dimensional (axisymmetric) simulations and includes the isotropic conduction case.
Here, instead of the ratio of cosmic-ray and plasma pressures (as in Fig. [11] of \cite{sha09}), we show metallicity. Like metals, 
cosmic rays (and possibly heat from other sources of ICM heating) are also more efficiently spread out (in both $r$ and $\theta$) with conduction than without it. It is curious that mixing with isotropic
thermal conduction is even more widespread than with anisotropic thermal conduction. These differences are explained in detail in the next section.

\section{Effect of Turbulent Forcing}
\begin{figure}
\psfrag{A}{$\Delta z$}
\psfrag{g}{$g$}
\psfrag{gT}{$\nabla T$}
\psfrag{Tb}{$T_b$}
\psfrag{T0}{$T_0$}
\psfrag{T1}{$T_0+\Delta z dT/dz$}
\includegraphics[width=4in,height=1in]{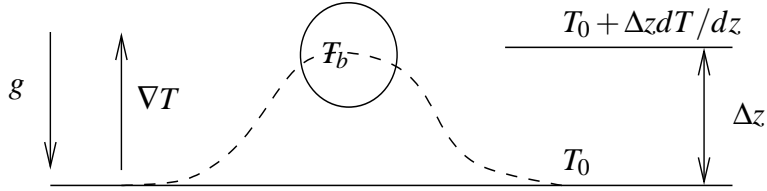}
\caption{Cartoon for the blob displaced from the initial stable/saturated HBI state (with horizontal field lines shown by the solid line) for $dT/dz>0$ ($\hat{z}$ is taken along the radial direction). The blob is displaced vertically by $\Delta z$; the dashed line shows the perturbed weak magnetic field. In the fast-conduction limit the blob temperature is the same as  the temperature of the initial field line; i.e., $T_b=T_0$. In the Boussinesq limit the blob pressure equals the background pressure at $\Delta z$ ($=p_0+\Delta zdp/dz$). The buoyancy force on the blob is $(\rho_b - \rho_0 - \Delta z d\rho/dz)g$, corresponding to the density difference of the blob relative to its surroundings. Expressing the density in terms of the temperature and pressure, the buoyancy force reduces to $\rho g \Delta z d\ln T/dz$ in the direction of gravity, i.e., a restoring force.}\label{fig1}
\end{figure}

There is a fundamental difference between convection in anisotropically conducting plasmas and the more well-known Schwarzschild convection that arises in adiabatic plasmas.
While the energy in Schwarzschild convection is mainly transported by fluid motions, it is transported by thermal conduction in anisotropically conducting plasmas (even when $dT/dz>0$ [$\hat{z}$ is along the radial direction], in which case field lines in the saturated state are aligned almost perpendicular to the temperature gradient).
In addition, the turbulent velocities in the saturated state in buoyantly unstable, anisotropically conducting plasmas are very small
(e.g., table 1 in \cite{par08} shows that the rms Mach numbers are $\lesssim 10^{-3}$ even when the vertical temperature gradient is large and the temperatures are fixed at the vertical boundaries; also \cite{sha08}). In contrast, in Schwarzschild convection the
turbulent velocities can be much larger, and turbulent velocities are larger for larger entropy gradients across the box.

From the above discussion (that the velocities are negligible with free convection [MTI/HBI] in anisotropically conducting plasma) we can consider a static saturated state for these instabilities with anisotropic thermal conduction. Figure (\ref{fig1}) considers a blob being perturbed from its HBI saturated state ($dT/dz >0$ in the background plasma) with horizontal field lines. The perturbed blob is at the same temperature as
the original field line ($T_0$). In the Boussinesq limit the blob is at the same pressure as the background pressure at the perturbed
position. The buoyancy force on the blob is a restoring force ($\rho g \Delta z d\ln T/dz$), bringing the blob back to its original position (see the figure caption). In this sense the HBI saturated state with horizontal field lines (and negligible velocities) 
is the stable state of an anisotropically conducting plasma with $dT/dz>0$. Analogous considerations for the MTI saturated state 
with vertical field lines when $dT/dz<0$, show that a vertically displaced blob experiences a similar restoring force ($\rho g \Delta z |d\ln T/dz|$).\footnote{The vertical field lines are not isothermal, but the 
conductive heat flux ${\bf Q}$ in the displaced blob satisfies $\nabla \cdot {\bf Q} = 0$;
see \cite{qua08} who invoke a similar argument for the destabilization of vertical field lines with $dT/dz>0$.} 
Thus, it is clear that a horizontally (vertically) aligned magnetic field with negligible velocity
is the saturated or nonlinearly {\em stable} state for the HBI (MTI). In other words, horizontal field lines (when $dT/dz>0$) and vertical field lines (when $dT/dz<0$) are the stable configurations of the system,  and the HBI and MTI are buoyancy instabilities that bring the system to this stable configuration (thus in this sense an anisotropically conducting plasma is analogous to a stably stratified adiabatic plasma, in that a perturbed blob experiences a restoring force). 
Only tiny velocities (ignoring magnetic tension and assuming that the conduction time is shorter than the Brunt-V\"ais\"al\"a time) bring the initially unstable system to its stable state (with reoriented field lines); after that there is no turbulent driving and the velocities decay with time (\cite{par08} \& \cite{par07}).
 This is entirely different from  convection in adiabatic plasmas where turbulent velocities are driven as long as the entropy gradients are 
sustained by the boundary conditions at the upper and lower boundaries of the convective region.

Although the buoyant restoring force in the HBI/MTI-saturated state resists distortions of the magnetic field, if an anisotropically conducting plasma is stirred sufficiently strongly to drive it away from its stable magnetic configuration and tangle up the magnetic field,
then the buoyancy
forces in this tangled-magnetic-field state could in principle add to (or subtract from) the turbulent driving
force. This effect, however, is not 
expected to change the turbulent velocities by more than a factor of order unity.

To understand why mixing is more effective with isotropic conduction than with anisotropic conduction in Figure (\ref{fig}), consider a blob
perturbed from its mean position by $\Delta z$; now, since the conductivity is large, the displaced blob has almost the same temperature as the background plasma at the same height. In the Boussinesq approximation the blob's internal pressure is the same as the external pressure at the same 
height. When the blob has the same pressure and almost the same temperature as its surroundings, there is almost no density
difference, and hence the buoyancy force is nearly zero. Because of this neutral buoyant response  it is much easier for a turbulence-driving mechanism such as cosmic-ray buoyancy or galaxy motions to
disperse a plasma with isotropic conduction than an adiabatic or anisotropically conducting plasma.

The ability of driven turbulence to overcome the stabilizing effect of anisotropic conduction (i.e., the 
tendency to pull back to horizontal [vertical] field lines with $dT/dz>0$ [<0]) can be quantified in terms of the Richardson number, a quantity often used in studies of forced turbulence in stably stratified systems \cite{tur73}. The Richardson number ($Ri$) is the ratio of the restoring buoyant force to
the random convective ($\rho {\bf{u}\cdot \nabla u}$) force, which can be thought of as driving a random walk of the displaced blob. The effect of the stabilizing influence (either for stably stratified adiabatic plasma or for anisotropically conducting plasma) can be overcome if the turbulent forcing
rate is larger than the rate of the stable buoyant response (i.e., the Brunt-V\"ais\"al\"a frequency for a stably stratified plasma, or $[g |d\ln T/dz|]^{1/2}$ for an anisotropically conducting plasma). This condition for effective turbulent mixing can be recast into the form  $Ri<Ri_c$, where $Ri_c$  is the critical  Richardson number below which the random turbulent forces can
overcome the effects of the stable buoyant response. ($Ri_c\approx 1/4$ is found in hydrodynamic studies of stably stratified systems \cite{tur73}.) The precise value of $Ri_c$ for anisotropically 
conducting plasmas, and its dependence on the initial field geometry, will be determined in future
with numerical simulations.

\section{Discussion}

On cluster core scales, where the thermal conduction time is shorter than the Brunt-V\"ais\"al\"a timescale, a hypothetical, isotropically conducting plasma would have $Ri \approx 0$ (a neutral buoyant response to external forcing; see previous section), and hence the efficient mixing seen in Figure (\ref{fig}). 
On the other hand, 
anisotropically conducting intracluster plasma naturally evolves to its stable saturated state (i.e., magnetic field lines perpendicular to $\nabla T$ when $dT/dr>0$), in which 
the gentle positive temperature gradient (as compared to a steep entropy gradient) results in a small stabilizing buoyant force.  Defining
$Ri \equiv g r (d\ln T/d\ln r)/u^2$ for the anisotropically conducting core, where $g$ is the gravitational
acceleration, $u$ is the typical forcing velocity and $r$ is the radius ($r$ is assumed to be the only scale in the problem but if the most effective turbulent driving is at a different scale this definition should be rescaled accordingly); for adiabatic plasma the temperature gradient is replaced by the entropy gradient. Thus, the Richardson number evaluated for typical cluster conditions is
\begin{equation}
Ri \approx 3 g_{-8} r_{10} \frac{d\ln T/d\ln r}{u_{100}^2},
\label{eq}
\end{equation}
where $g_{-8}$ is the gravitational acceleration in the units of $10^{-8}$ cm$^{2}$s$^{-1}$, $r_{10}$ is the
scale height (taken roughly to be the radius) in the units of 10 kpc, $u_{100}$ is the shear velocity in the units of 100 km s$^{-1}$. A typical logarithmic temperature gradient in cluster cores is $d\ln T/d \ln r \approx$ 0.15,  and the typical logarithmic entropy gradient is $d\ln s/d\ln r \approx 0.6$ (e.g., \cite{pif05}). Thus, from the requirement that $Ri < Ri_c$ (since only a single spatial scale is used in defining $Ri$ in Eq. [\ref{eq}],  $Ri_c \approx 1/4$ is only a rough estimate for the critical value) for efficient turbulent mixing, 
turbulent velocities $\sim$ 100 km s$^{-1}$ are sufficient for efficient mixing in the 
cluster core and for overcoming the restoring force experienced by a plasma in the HBI-saturated state
(see Fig. [5] in \cite{sha09}; similar velocities are seen in vigorously stirred regions in the top panel of Fig. [\ref{fig}]; although turbulent velocities in \cite{sha09} are driven by cosmic-ray convection the above criterion for turbulent mixing applies more generally). Because 
$ds/d\ln r \simeq 4 |d\ln T/dr|$, a four-times-larger turbulent
energy would be required to induce forced mixing in a hypothetical adiabatic cluster core. Since isotropically conducting plasma has a neutral buoyant response, it can be effectively mixed by much smaller turbulent velocities.
Consistent with the above, the turbulent velocities (in the well-mixed regions) in the isotropically conducting case are the smallest, followed by the anisotropically conducting case, and by the adiabatic case.

Although these considerations provide a qualitative understanding of mixing and turbulence in the ICM, realistic numerical simulations, including cooling, are required for understanding the role of forced turbulence (e.g., by cosmic rays, shear instabilities at the jet boundary, etc.) in heating the ICM and spreading the heat isotropically.
Although we have focused on cluster cores with $dT/dr>0$, similar considerations for mixing and turbulence apply at larger radii where $dT/dr<0$. The ICM plasma can be driven turbulent by the infalling dark-matter halos and galaxy wakes if large turbulent velocities (10-100 km/s) are generated. The stability of the anisotropic plasma in its saturated state (i.e., horizontal field lines with $dT/dr>0$ and vertical field lines with $dT/dr<0$) has important implications for the effective thermal conductivity of the ICM. It is sometimes argued (e.g., \cite{nar01}) that the ICM should be turbulent and the effective thermal conductivity should be a fraction ($\approx 1/3$) of the Spitzer value. In light of the arguments presented in this paper, isotropic turbulence (leading to  $\approx 1/3$ Spitzer conduction) 
can only be driven if the magnitude of turbulent driving is reasonably large ($\sim$10-100 km s$^{-1}$); for a lower magnitude of turbulent stirring the ICM plasma will relax to 
its stable state with anisotropic thermal conduction (i.e., horizontal field lines in the ICM core where $dT/dr>0$ and vertical field lines in outer regions where $dT/dr<0$). Thus, in the absence of vigorous turbulent stirring, the effective conductivity (normalized to the Spitzer value) is $\lesssim 0.1$ (e.g., \cite{par08}) in the ICM core and $\sim 1$ in the outer regions. Such a drastic modification of the thermal conductivity has implications for the role of thermal conduction in solving the cooling flow problem (e.g., see \cite{par09}).

The stability of a stably stratified adiabatic plasma  and an anisotropically conductive plasma is analogous to the hydrodynamic stability of Keplerian flows (e.g., \cite{haw99}).
In both cases there is a restoring force (epicyclic stabilization for Keplerian flows and buoyant stabilization in anisotropically conducting stratified plasmas) that brings the 
perturbed fluid element back to its original location. The focus in hydrodynamic disks has been on the nonlinear stability of Keplerian flows and not so much on the forced 
driving of turbulence and transport, but it is likely that a Richardson-like criterion, that the shearing rate due to the sustained nonlinear velocities be larger than the 
epicyclic frequency, holds for these linearly stable hydrodynamic shear flows.

In addition to the effect of thermal conduction on free convection (via HBI, MTI),
its effect on forced convection (e.g., turbulent  forcing by jets, cosmic rays, etc.) is also crucially important, especially in cluster cores, where strong kinetic feedback mechanisms are expected to operate.  In most of the discussion, the effect of magnetic tension is ignored and the conduction
time is assumed to be much shorter than buoyancy timescale.
For more details on the numerical setup, initial and boundary conditions, etc. refer to \cite{sha09}.

\begin{theacknowledgments}
Support for this work was provided by NASA through Chandra Postdoctoral Fellowship grant numbers PF8-90054 (to PS, EQ) and 
PF7-80049 (to IJP, EQ) awarded by the Chandra X-ray Center, which is 
operated by the Smithsonian Astrophysical Observatory for NASA under contract NAS8-03060.  This research was 
supported in part by the National Science Foundation through 
TeraGrid resources provided by NCSA and Purdue University.
\end{theacknowledgments}

\end{document}